\documentclass[aps,prl,floatfix,twocolumn]{revtex4-1}
\usepackage{times}
\usepackage{graphicx}
\usepackage{amsmath}
\usepackage{amssymb}
\usepackage{textcomp} 

\begin{document}
\title{A solver based on pseudo-spectral analytical time-domain method for the two-fluid plasma model}

\author{B. Morel$^{\ast}$, R. Giust, K. Ardaneh and F Courvoisier$^{\ast}$\\
FEMTO-ST institute, Univ. Bourgogne Franche-Comt\'e, CNRS,\\
15B avenue des Montboucons, 25030, Besan\c{c}on Cedex, France\\
$^{\ast}$ Corresponding authors: benoit.morel@femto-st.fr and  francois.courvoisier@femto-st.fr\\
\vspace{2cm}
This is a pre-peer-review version of an article published in {\it Scientific Reports}. The final version is available online at: \\
\url{https://doi.org/10.1038/s41598-021-82173-9}
\vspace{2cm}
}

\begin{abstract}
A number of physical processes in laser-plasma interaction can be described with the two-fluid plasma model. We report on a solver for the three-dimensional two-fluid plasma model equations. This solver is particularly suited for simulating the interaction between short laser pulses  with plasmas. The fluid solver relies on two-step Lax-Wendroff split with a fourth-order Runge-Kutta scheme, and we use the PseudoSpectral Analytical Time-Domain (PSATD) method to solve Maxwell's curl equations. Overall, this method is only based on finite difference schemes and fast Fourier transforms and does not require any grid staggering. The PseudoSpectral Analytical Time-Domain method removes the numerical dispersion for transverse electromagnetic wave propagation in the absence of current that is conventionally observed for other Maxwell solvers. The full algorithm is validated by conservation of energy and momentum when an electromagnetic pulse is launched onto a plasma ramp and by quantitative agreement with wave conversion of p-polarized electromagnetic wave onto a plasma ramp.
\end{abstract}
\maketitle


\section*{Introduction}

Chirped pulse amplification technology in 1985 \cite{Strickland_1985} has made possible the generation of extremely powerful laser pulses \cite{eliezer_applications_2008}. When a solid, liquid or a gas is irradiated by such a powerful pulse, the ionization phenomena swiftly create a plasma at the surface of the material or within the gas. The development of applications such as inertial fusion \cite{Basov_1964}, laser-plasma accelerators \cite{Esarey_1996}, laser materials processing \cite{eliezer_applications_2008}, X-Ray lasers \cite{Daido_2002} or nonlinear plasmonics at lower intensities\cite{Kauranen_2012} requires laser plasma interactions modeling.

Hydrodynamic models are particularly useful to describe short pulse interaction with plasmas when each of the species can be assumed in local thermodynamic equilibrium \cite{mckenna_laser-plasma_nodate}. The two-fluid plasma equations is the starting point of the hydrodynamic models \cite{gibbon_short_2005}. This model describes the spatio-temporal evolution of the density, mean velocity and pressure of electrons and ions fluids. The two-fluid plasma equations therefore consist of two sets of Euler equations with source term, as well as Maxwell's equations. The fluid description involves the assumptions of local thermodynamic equilibrium for each species (electrons, ions). The conventional hydrodynamic models, {\it e.g.}, two-temperature plasma equations, single-fluid equations and MagnetoHydroDynamic (MHD) can be derived from the two-fluid plasma model by means of additional assumptions.

At present, solving the complete two-fluid plasma equations is a difficult  challenge\cite{abgrall_robust_2014}. Their implementation is often complex for non-specialist groups since most of these codes are developed to be particularly robust for shock's problems. A good example is given in Shumlak {\it et al.} \cite{shumlak_approximate_2003} which present an algorithm based on Roe-type Riemann solver \cite{roe_approximate_1981} for the two-fluid plasma model. The same group added the high-order discontinuous Galerkin method to improve the result's accuracy \cite{loverich_discontinuous_2005,loverich_discontinuous_2011,srinivasan_analytical_2011, sousa_blended_2016}. References \cite{alvarez_laguna_fully-implicit_2018,mason_electromagnetic_1987,mason_hybrid_1986,baboolal_two-scale_2007,kumar_entropy_2012} describe numerical methods well adapted for shock's problems. In contrast, for problems without strong shocks, our group has recently proposed a relatively simple approach \cite{Morel_2020}, based on finite difference schemes and Fast Fourier Transform (FFT). This solver is based on the PseudoSpectral Time-Domain method (PSTD) \cite{liu_pstd_1997} to solve Maxwell equations and a composite scheme \cite{liska_composite_1998} to solve the fluid equations. However, the PSTD is based on a temporally staggered grid, which requires temporal interpolations for the coupling with the fluid solver. In addition, the PSTD algorithm is numerically dispersive. It emits wave components that are faster than light.

J.-L. Vay {\it et al.} \cite{Vay_2013} proposed the PseudoSpectral Analytical Time-Domain (PSATD), and its application to pseudo-spectral Particle-In-Cell (PIC) simulations. In PSTD, the temporal integration is performed via finite differences, while in PSATD, the integration is analytical (except for the integration of current). Thus, unlike PSTD, the PSATD requires no temporal staggered grid and is free of numerical dispersion for transverse electromagnetic propagation in the absence of current (see Fig. 1 of J.L. Vay \textit{et al.} \cite{Vay_2013}). This method is also particularly well adapted for laser pulse propagation. The algorithm is tested with laser/plasma interaction problems with intensities around $10^{14}$ to $10^{15}$~W/cm$^2$, since it is intended for the study of electron/hole plasma dynamics in solids. 

\begin{figure}[h!]
\centering
  \includegraphics[width=\columnwidth]{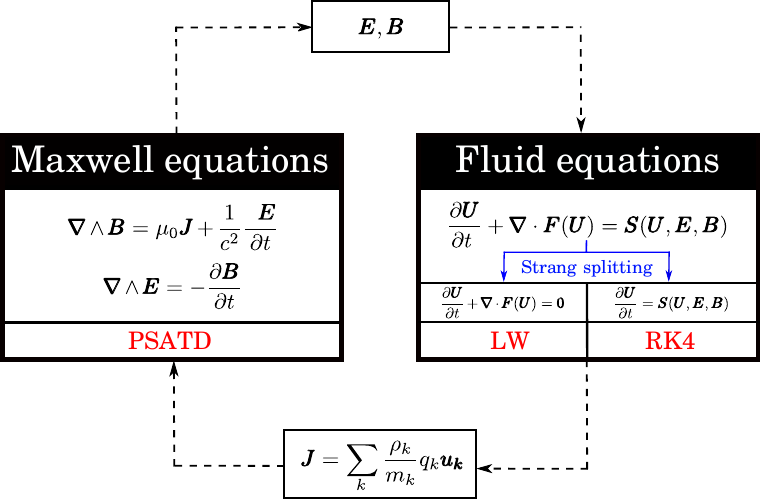}
  \caption{Schematic representation of the solver's structure.}
  \label{fig:Ov1}
\end{figure}

Here, we build a two-fluid plasma solver based on PseudoSpectral {\it Analytical} Time-Domain PS{\it A}TD for solving Maxwell's equations. A schematic representation of our solver is given in Fig. \ref{fig:Ov1}. The integration of Maxwell's equations is performed by using the PSATD method. The electromagnetic fields are transmitted to the fluid equations as a Lorentz force source term. The fluid equations are integrated by using a Strang splitting \cite{strang_construction_1968}. In the splitting, the homogeneous system is solved via a Lax Wendroff (LW) scheme, while the source terms are integrated with a fourth-order Runge-Kutta scheme (RK4). The updated fluid variables are used to calculate the current density, which is injected in Maxwell's equations.

This paper is divided in four main parts. We will first recall the two-fluid plasma model equations before summarizing the numerical integration. We will then validate the solver and demonstrate its benefits in terms of numerical dispersion and in the reduction of the constraint imposed on time-steps with the solver of reference \cite{Morel_2020}.

\section*{Results}

\subsection*{Two-fluid plasma equations}

The two-fluid plasma model equations consists of Euler equations with source term for each fluid, as well as Maxwell's equations. This system of equations corresponds to continuity equations, motion equations and energy transport equations for electron and ion fluids. Following reference \cite{a._hakim_high_2006}, the fluid equations can be presented under the following form:
\begin{equation}
\frac{\partial }{\partial t} \underbrace{ \left[ \begin{matrix}
\rho_\mathrm{e} \\
\rho_\mathrm{e} \mathbf{u_\mathrm{e}} \\
\epsilon_\mathrm{e} \\
\rho_\mathrm{i} \\
\rho_\mathrm{i} \mathbf{u_\mathrm{i}} \\
\epsilon_\mathrm{i} \\
\end{matrix} \right]}_{\equiv \mathbf{U}}   + \mathbf{\nabla} \cdot \underbrace{\left[ \begin{matrix}
\rho_\mathrm{i} \mathbf{u_\mathrm{e}} \\
\rho_\mathrm{e} \mathbf{u_\mathrm{e}} \otimes \mathbf{u_\mathrm{e}} + p_\mathrm{e} \mathbf{I}   \\
(\epsilon_\mathrm{e} + p_\mathrm{e}) \mathbf{u_\mathrm{e}} \\
\rho_\mathrm{i} \mathbf{u_\mathrm{i}} \\
\rho_\mathrm{i} \mathbf{u_\mathrm{i}} \otimes \mathbf{u_\mathrm{i}} + p_\mathrm{i} \mathbf{I} \\
(\epsilon_\mathrm{i} + p_\mathrm{i}) \mathbf{u_\mathrm{i}}
\end{matrix} \right]}_{\equiv \mathbf{F}(\mathbf{U})} = \underbrace{\left[ \begin{matrix}
0 \\
\frac{\rho_\mathrm{e} q_\mathrm{e}}{m_\mathrm{e}} (\mathbf{E} + \mathbf{u_\mathrm{e}} \mathbf{\times} \mathbf{B} ) \\
\frac{\rho_\mathrm{e} q_\mathrm{e}}{m_\mathrm{e}} \mathbf{u_\mathrm{e}} \cdot \mathbf{E}  \\
0 \\
\frac{\rho_\mathrm{i} q_\mathrm{i}}{m_\mathrm{i}} (\mathbf{E} + \mathbf{u_\mathrm{i}} \mathbf{\times} \mathbf{B} ) \\
\frac{\rho_\mathrm{i} q_\mathrm{i}}{m_\mathrm{i}} \mathbf{u_\mathrm{i}} \cdot \mathbf{E} 
\end{matrix}  \right]}_{\equiv \mathbf{S}(\mathbf{U},\mathbf{E},\mathbf{B})} 
\label{Econs}
\end{equation}    
where $\mathbf{U}$ is the fluid variables vector, $\mathbf{F}(\mathbf{U})$ is the flux tensor and $\mathbf{S}(\mathbf{U},\mathbf{E},\mathbf{B})$ is the Lorentz force source term.
In this paper, $i$ and $e$ are indexes related respectively to the ion fluid and to the electron fluid. $q$ is the charge, $m$ the mass, $\rho$ the mass density, $\mathbf{u}$ the mean velocity, $p$ the pressure, $\epsilon$ the fluid energy density, $\mathbf{E}$ the electric field and $\mathbf{B}$ the magnetic field. $\otimes$ is tensor product and $\mathbf{I}$ is the identity matrix. 

One more equation is required to close the system of equation, an ideal gas closure for each fluid $k$ is used \cite{a._hakim_high_2006}:
\begin{equation}
\epsilon_\mathrm{k} \equiv \frac{p_\mathrm{k}}{\gamma -1 } +  \frac{1}{2}\rho_\mathrm{k} \mathbf{u_\mathrm{k}}^2
\label{Eq_Energy}
\end{equation}
where $\gamma$ is the adiabatic index. 

 Electric and magnetic fields in the source term $\mathbf{S}(\mathbf{U},\mathbf{E},\mathbf{B})$ are determined by the Maxwell equations. Since solving fluid equations and Maxwell's curl equations enforces the conservation of divergence properties of the fields \cite{bittencourt_fundamentals_2013}, it is not necessary to solve Maxwell's divergence equations. However they must be satisfied at an initial time. Furthermore, Maxwell curl's equations can be written by expressing the current density $\mathbf{J}$ as function of fluid variables:
\begin{equation}
\mathbf{\nabla} \mathbf{\times} \mathbf{E} = - \frac{\partial \mathbf{B}}{\partial t}
\label{curl1}
\end{equation}
 \begin{equation}
\mathbf{\nabla} \mathbf{\times} \mathbf{B} = \mu_\mathrm{0}  \underbrace{\left[ \frac{q_\mathrm{e}}{m_\mathrm{e}}\rho_\mathrm{e} \mathbf{u_\mathrm{e}} + \frac{q_\mathrm{i}}{m_i}\rho_\mathrm{i} \mathbf{u_\mathrm{i}} \right]}_{\textnormal{$\equiv \mathbf{J}$}}  + \frac{\varepsilon_\mathrm{r}}{c^2}\frac{\partial \mathbf{E}}{\partial t}
\label{MA}
\end{equation}
Here, $\varepsilon_\mathrm{0}$ and $\mu_\mathrm{0}$ are respectively the vacuum permittivity and permeability. $c=(\varepsilon_\mathrm{0} \mu_\mathrm{0})^{-1/2}$ is the speed of light. $\varepsilon_\mathrm{r}$ is the relative permittivity of the background medium: we do the assumption that this quantity is time and space independent. In this model, the plasma is also contained inside a medium of relative permittivity $\varepsilon_\mathrm{r}$.

\subsection*{The numerical integration}

\subsubsection*{The Maxwell solver}

As mentioned in the introduction, the PSATD method \cite{Vay_2013} is used for solving Maxwell curl's equations. This method is simple to implement and does not need the staggering of spatial and temporal grids. This is in contrast with the Finite Difference Time Domain (FDTD) method \cite{yee_numerical_1966} which requires spatially and temporal staggered grids or in contrast with the PseudoSpectral Time-Domain (PSTD) \cite{liu_pstd_1997} which requires a temporally staggered grid. The PSATD is therefore more flexible to be coupled with another algorithm without interpolations. Moreover, in absence of current, PSATD induces zero numerical dispersion in contrast with FDTD or PSTD.
An additional strong benefit is that the PSATD is not subject to a Courant condition for transverse electromagnetic field propagation in the absence of current. 
The PSATD algorithm is inherently periodic because it is based on FFT, but open systems can be modeled by using Perfectly Matched Layers (PML) as in O. Shapoval \textit{et al.}\cite{Shapoval_2019}.

The PSATD algorithm provides the fields in the Fourier space \cite{Vay_2013}:
\begin{equation}
\label{Eq_PSATD1}
\begin{split}
     \mathbf{\tilde{E}}^\mathrm{n+1} = C_\mathrm{0} \mathbf{\tilde{E}}^\mathrm{n} + i v S_\mathrm{0} \boldsymbol{\kappa} \times \mathbf{\tilde{B}}^\mathrm{n}  -  \frac{1}{\varepsilon_0 \varepsilon_r} \frac{S_\mathrm{0}}{kv}  \mathbf{\tilde{J}}^\mathrm{n+1/2} \\+ (1-C_\mathrm{0}) \boldsymbol{\kappa} \cdot (\boldsymbol{\kappa} \cdot \mathbf{\tilde{E}}^\mathrm{n})  + \frac{1}{\varepsilon_0 \varepsilon_r } \left( \frac{S_\mathrm{0}}{kv}-\Delta t \right) \boldsymbol{\kappa} \cdot (\boldsymbol{\kappa} \cdot \mathbf{\tilde{J}}^\mathrm{n+1/2}) 
		\end{split}
\end{equation}
\begin{equation}
     \mathbf{\tilde{B}}^\mathrm{n+1} = C_\mathrm{0} \mathbf{\tilde{B}}^\mathrm{n} - i \frac{S_\mathrm{0}}{v} \boldsymbol{\kappa} \times \mathbf{\tilde{E}}^\mathrm{n}   +i  \mu_0 \frac{1-C_\mathrm{0}}{k} \boldsymbol{\kappa} \times \mathbf{\tilde{J}}^\mathrm{n+1/2}
     \label{Eq_PSATD2}
\end{equation}
where $\mathbf{\tilde{a}}$ is the Fourier transform of the quantity $\mathbf{a}$. Here $C_\mathrm{0}=\cos \left(k v \Delta t\right)$, $S_\mathrm{0}=\sin \left(k v \Delta t\right)$, $ \boldsymbol{\kappa}=  \mathbf{k}/k$ and $v=\frac{c}{\varepsilon_r^{1/2}}$. The two main assumptions made in the PSATD method are: 1) the time-step $\Delta t$ is enough small to assume that the current density is constant over a time-step 2) the background permittivity $\varepsilon_\mathrm{r}$ is uniform.

\subsubsection*{The fluid solver}

For the fluid equations solver, we consider a similar solver as in reference \cite{Morel_2020}. Here, we simplified the solver by restricting ourselves to problems without discontinuities such that 
it becomes unnecessary to introduce numerical dissipation to make gradient smoother. Instead of using a composite scheme LWLFn as in reference \cite{Morel_2020}, we will use a simple two-step Lax-Wendroff (LW) scheme \cite{richtmyer_survey_1962} which is second order accurate and introduces less numerical dissipation than the two-step Lax-Friedrichs (LF) scheme \cite{shampine_two-step_2005}.
The LW scheme solves the homogeneous part of Eq. \eqref{Econs}, as we recall below.

First, we set $L^\mathrm{x}$ the operator for the two-step LW along $x$ direction:
\begin{equation}
\begin{split}
L^\mathrm{x} (\mathbf{U}^\mathrm{n}_\mathrm{j,l,m} ) =  \mathbf{U}^\mathrm{n}_\mathrm{j,l,m} - \frac{\Delta t}{ \Delta x} \\ \times\left[ \mathbf{F_x} \left(\mathbf{U}^\mathrm{n+1/2}_\mathrm{j+1/2,l,m}\right) - \mathbf{F_x}\left(\mathbf{U}^\mathrm{n+1/2}_\mathrm{j-1/2,l,m}\right) \right] 
\end{split}
\label{SYS1}
\end{equation}
with
\begin{equation}
\begin{split}
\mathbf{U}^\mathrm{n+1/2}_\mathrm{j+1/2,l,m} = \frac{1}{2} \left[ \mathbf{U}^\mathrm{n}_\mathrm{j+1,l,m} + \mathbf{U}^\mathrm{n}_\mathrm{j,l,m} \right] \\
- \frac{\Delta t}{2 \Delta x} \left[ \mathbf{F_x}\left(\mathbf{U}^\mathrm{n}_\mathrm{j+1,l,m}\right) - \mathbf{F_x}\left(\mathbf{U}^\mathrm{n}_\mathrm{j,l,m}\right) \right]
\end{split}
\end{equation}
\begin{equation}
\begin{split}
\mathbf{U}^\mathrm{n+1/2}_\mathrm{j-1/2,l,m} = \frac{1}{2} \left[ \mathbf{U}^\mathrm{n}_\mathrm{j,l,m} + \mathbf{U}^\mathrm{n}_\mathrm{j-1,l,m} \right] \\
- \frac{\Delta t}{2 \Delta x} \left[ \mathbf{F_x}\left(\mathbf{U}^\mathrm{n}_\mathrm{j,l,m}\right) - \mathbf{F_x}\left(\mathbf{U}^\mathrm{n}_\mathrm{j-1,l,m}\right) \right]
\end{split}
\end{equation}
where $j$, $l$ and $m$ are respectively indexes for $x$, $y$ and $z$ directions. Similar operations are done in $y$ and $z$ directions as reference \cite{Morel_2020} to obtain $L^\mathrm{y} (\mathbf{U}^\mathrm{n}_\mathrm{j,l,m})$ and $L^\mathrm{z} (\mathbf{U}^\mathrm{n}_\mathrm{j,l,m})$. A basic spatially dimensionally-split scheme is used to obtain the value $\mathbf{U}^\mathrm{n+1}_\mathrm{j,l,m}$ from $\mathbf{U}^\mathrm{n}_\mathrm{j,l,m}$ \cite{leveque_finite_2002}:
\begin{equation}
\mathbf{U}^\mathrm{n+1}_\mathrm{j,l,m} =   L^\mathrm{x} L^\mathrm{y} L^\mathrm{z} ( \mathbf{U}^\mathrm{n}_\mathrm{j,l,m} )
\label{Symmetrization}
\end{equation}

For the numerical integration of the source term $\mathbf{S}(\mathbf{U},\mathbf{E},\mathbf{B})$ of Eq. \eqref{Econs}, we use the Strang splitting technique presented by G. Strang\cite{strang_construction_1968}. The Strang splitting allows an estimation of current density $\mathbf{J}^\mathrm{n+1/2}$ at a half time step of PSATD. The concept of Strang splitting is shown on the steps 1, 2 and 4 in Fig. \ref{fig:Algo}. We first integrate the source term with an RK4 scheme over $\Delta t /2$, then the homogeneous system is integrated over $\Delta t$ with an LW scheme, and finally source term is again integrated with an RK4 over time step $\Delta t /2$.

\subsubsection*{Full two-fluid plasma solver algorithm}

\begin{figure*}[h]
\centering
\includegraphics[scale=0.75]{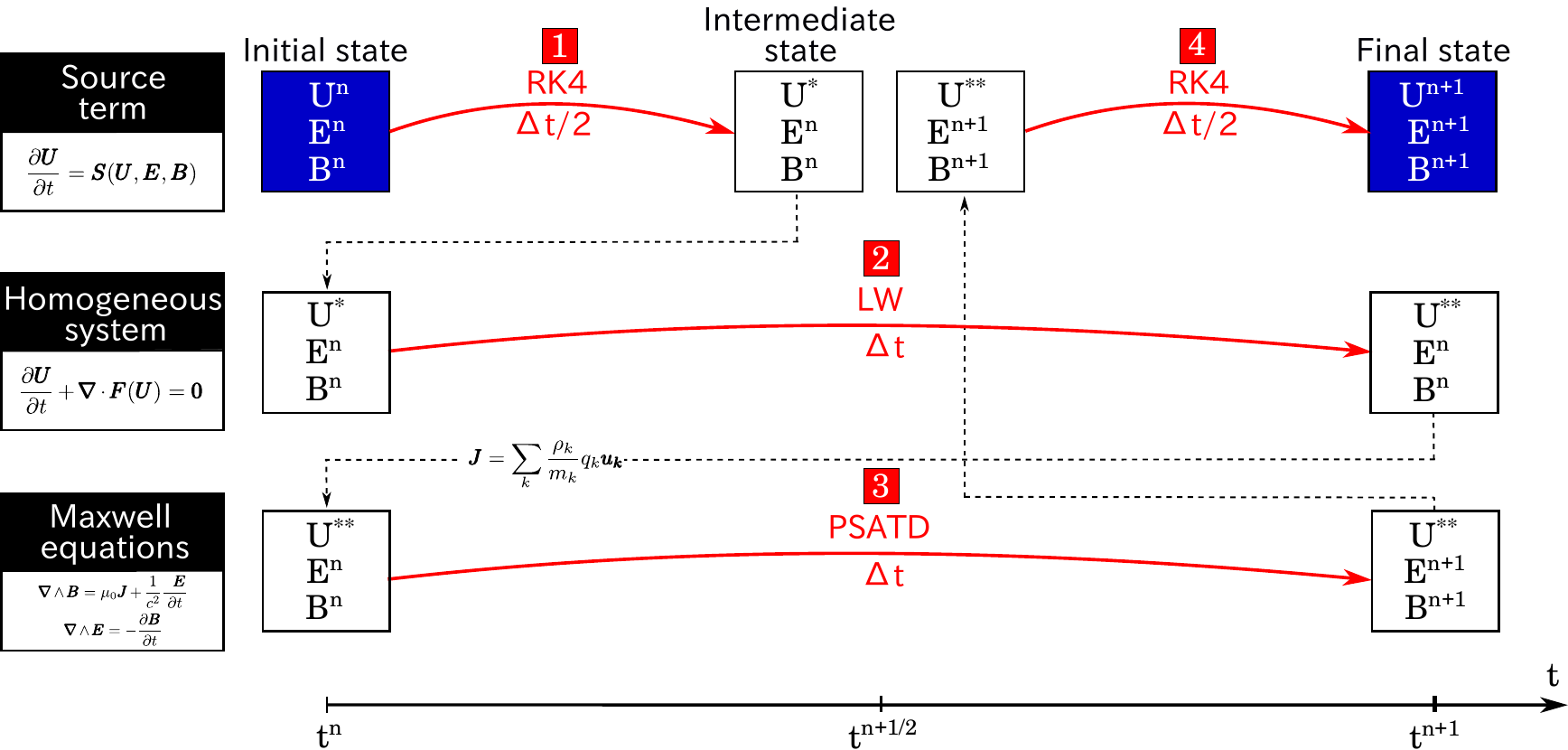}
  \caption{Schematic representation of the algorithm. The source term is integrated with an RK4 scheme while the homogeneous system is integrated with an LW scheme. Maxwell's equations are solved with the PSATD method. The algorithm requires four steps shown in red to advance fluid variables and fields from a time step $n$ to $n+1$.}
  \label{fig:Algo}
\end{figure*}

The full algorithm for the two-fluid plasma model is described in Fig. \ref{fig:Algo} and can be decomposed in 4 main steps:
\begin{enumerate} 
\item Integration of the source term with an RK4 scheme over a temporal step $\Delta t/2$ by using $\mathbf{E}^\mathrm{n}$, $\mathbf{B}^\mathrm{n}$ and $\mathbf{U}^\mathrm{n}$ to obtain the intermediate value of fluid variables $\mathbf{U^*}$.
\item Integration of the homogeneous system with an LW scheme over a temporal step $\Delta t$ using fluid variables vector $\mathbf{U}^{*}$ to obtain a new intermediate value $\mathbf{U}^{**}$.
\item Computation of the current density $\mathbf{J}^\mathrm{n+1/2}$ with densities and velocities from $\mathbf{U}^{**}$. Then, carry out a PSATD step with $\mathbf{J}^\mathrm{n+1/2}$ to calculate $\mathbf{E}^\mathrm{n+1}$ and $\mathbf{B}^{\mathrm{n+1}}$. 
\item Integration of the source term with an RK4 algorithm over a temporal step $\Delta t/2$ using $\mathbf{U}^{**}$, $\mathbf{E}^\mathrm{n+1}$ and $\mathbf{B}^\mathrm{n+1}$ to obtain the final value of fluid variables $\mathbf{U}^\mathrm{n+1}$.
\end{enumerate} 
The PSATD naturally represents all field values at the nodes of a grid, it also avoids temporal interpolation of the magnetic field that was necessary in reference \cite{Morel_2020}.

For the PSATD algorithm alone without currents, the sampling is in principle only limited by Nyquist theorem. However, in order to derive Eqs. \eqref{Eq_PSATD1} and \eqref{Eq_PSATD2}, we make the assumption that the current is constant over the temporal step $\Delta t$. Therefore, the temporal step is chosen small enough to make this assumption valid. The spatial step $\Delta x$ is simply chosen to resolve the both plasma and electromagnetic waves.

\subsection*{Validation of the numerical solver}

\subsubsection*{S-polarized electromagnetic wave over a plasma ramp}

\begin{figure*}[h!]
\centering
  \includegraphics[width=\textwidth]{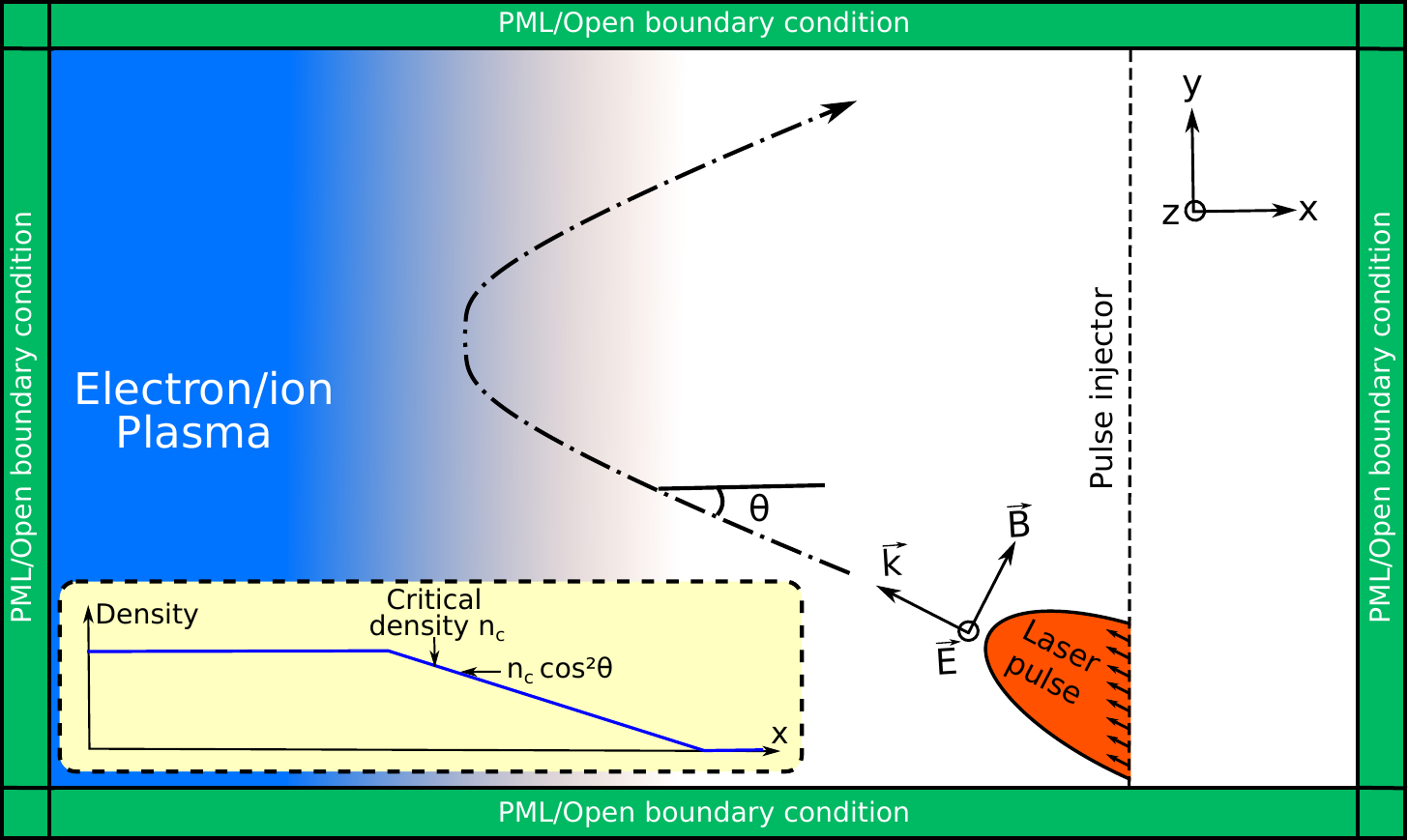}
  \caption{Numerical setup: A laser pulse in oblique incidence is injecting toward an overcritical plasma ramp. The laser pulse is reflected at the turning surface such as density $n=n_c \cos^2 \theta$.}
  \label{fig:Box}
\end{figure*}

In this first test, we check the conservation of momentum and energy during reflection of a s-polarized electromagnetic wave over a plasma ramp. The numerical setup is shown in Fig. \ref{fig:Box}. A laser pulse is propagating toward an overcritical plasma ramp with an angle of incidence $\theta =15^\circ$. The initial plasma density profile is invariant in $y$ and $z$ directions, and the following initial density profiles, for electron and ion fluids, are used in $x$ direction:
\begin{equation}
n_\mathrm{e} = n_\mathrm{i} = \left\lbrace
 \begin{array}{ccc}
0 &  \textnormal{for } x>-5\mu m\\
 0.57 \times 10^{21} (5-x) & \textnormal{for }  -5\mu m\leq x \leq -10\mu m\\
2.85 \times 10^{21} & \textnormal{for } x < -10\mu m
\end{array}\right.
\end{equation}
where $n_\mathrm{e}\equiv \rho_\mathrm{e}/m_\mathrm{e}$ and $n_\mathrm{i}\equiv \rho_\mathrm{i}/m_\mathrm{i}$ are given in cm$^\mathrm{-3}$. The length in $x$ direction at which the critical density $n_\mathrm{c} = 1.75 \times 10^\mathrm{21}$~cm$^\mathrm{-3}$ is reached is $L=3.08~\mu$m. We add a weak uniform background density of $10^{17}$~cm$^{-3}$ to avoid divisions by zero in the algorithm and too strong discontinuity at the ramp onset. In this test, the uniform background is vacuum: $\varepsilon_r=1$. For the plasma, we take $m_\mathrm{e}= 9.11 \times 10^{-31}$~kg, $m_\mathrm{i}=1837m_\mathrm{e}$ and $\gamma=5/3$. The initial mean velocities and pressure are zero.

The laser pulse is a spatially Gaussian beam with a waist $w_0 = 4~\mu$m and is described temporally by a single period of a $\sin^2$ function (period $T = 40$~fs). The free-space wavelength is $\lambda = 0.8~\mu$m and the amplitude in free-space is $E_{0} = 4.3 \times 10^{10}$~V/m. We choose this electric field amplitude to demonstrate the possibility of working with high field amplitudes with this algorithm. Note that the beam is invariant along $z$ direction. 

The number of points in $x$ and $y$ directions is $N_\mathrm{x}=N_\mathrm{y}=512$ and $N_\mathrm{z}=2$ is $z$ direction. PML (resp. open) boundary conditions in $x$ and $y$ directions for PSATD (resp. fluid algorithm) are implemented. For fluids and fields, periodic boundary condition are used in $z$ direction. The spatial step is $\Delta x = \Delta y= \Delta z= 60 $~nm and the temporal step is $\Delta = 96$~as.

\begin{figure*}[h!]
\centering
  \includegraphics[width=\textwidth]{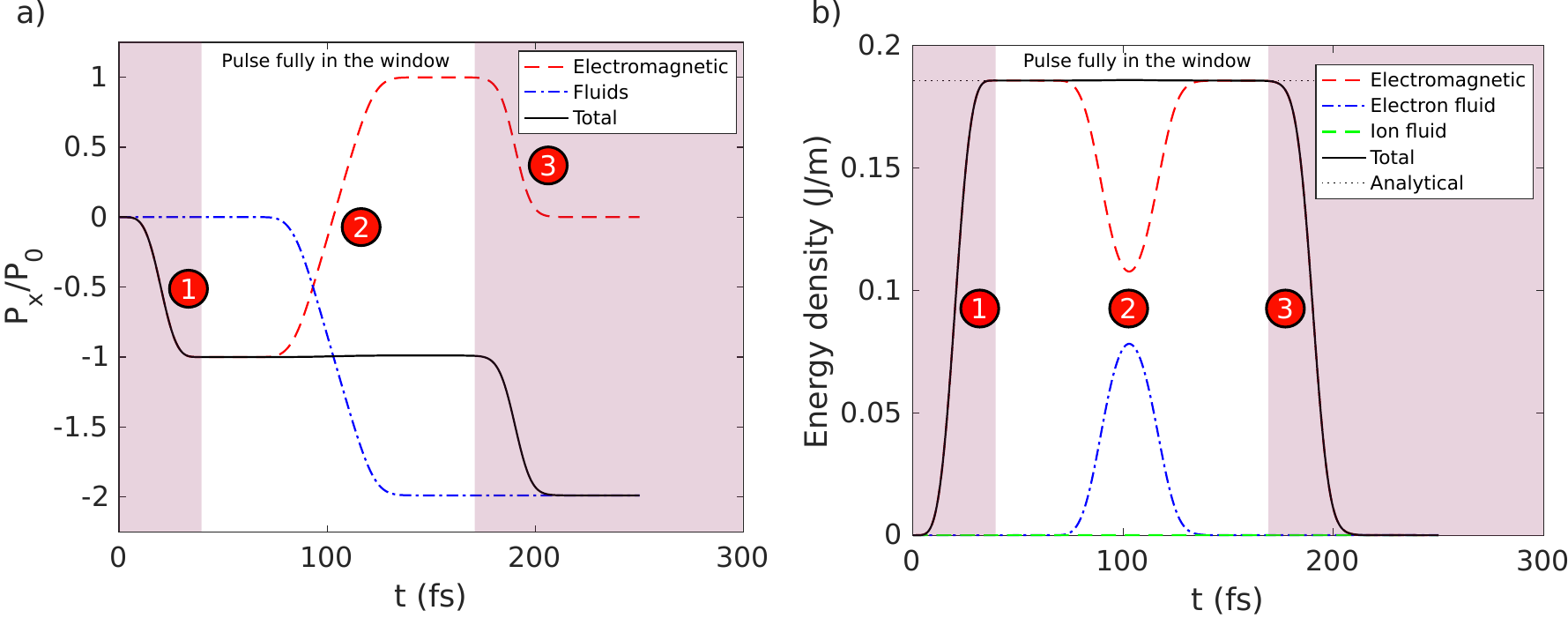}
  \caption{(a) Conservation of momentum and (b) conservation of energy during s-polarized pulse reflection over the plasma ramp. In the central white area, the pulse is fully in the numerical window and conservation of momentum and energy are preserved.  }
  \label{fig:Conservation}
\end{figure*}

 In Fig. \ref{fig:Conservation}(a), we plot the different momenta in $x$ direction as function of the simulation time. These momenta are normalized to the absolute value of the $x$ momentum $P_\mathrm{0}$ of the incident pulse. To measure the normalization factor $P_\mathrm{0}$, we performed beforehand the simulation without the plasma, and we measure the $x$ momentum of the pulse defined by the $x$ component of the Eq. \eqref{Eq_Moment} integrated over the simulation window.
 
 The density of electromagnetic momentum is defined by \cite{Jackson_1999}:
\begin{equation}
\mathbf{P}_\mathrm{em}= \varepsilon_0 \mathbf{E} \times \mathbf{B}
 \label{Eq_Moment}
\end{equation}
The dashed red curve of Fig. \ref{fig:Conservation}(a) corresponds to the normalized electromagnetic $x$ momentum density integrated in the simulation window. The dashed dotted blue curve corresponds to the normalized fluids $x$ momentum integrated in the simulation window. The momentum of fluids is defined by:
 \begin{equation}
    \mathbf{P_\mathrm{f}} = \rho_\mathrm{e} \mathbf{u}_\mathrm{e} + \rho_\mathrm{i} \mathbf{u}_\mathrm{i} 
 \end{equation}
 The black line of Fig. \ref{fig:Conservation}(a) is the sum of the electromagnetic and fluids $x$ momentum.

 We observe three main sequences in Fig. \ref{fig:Conservation}(a):
 \begin{itemize}
     \item \textbf{1}: Since the laser pulse goes from the right to the left, the electromagnetic momentum along $x$ (red dashed curve) decreases as the pulse enters into the simulation window (between $t=0$ and $t<50$~fs). At $t=50$~fs, the pulse is completely contained in the simulation window and has not yet interacted with the plasma ramp. We see that the electromagnetic $x$ momentum corresponds to the incident pulse $x$ momentum $-P_{0}$.
     \item \textbf{2}: In the temporal window 70-130~fs, momentum exchange with the plasma takes place: the fluids momentum decreases until $-2P_0$, whereas the electromagnetic $x$ momentum increases until reaches $+P_0$. This is the signature that the laser pulse transfers twice its initial momentum to the plasma during its reflection, as can be expected.
     \item \textbf{3}: Between $170$~fs and $210$~fs, the reflected pulse leaves the simulation window, thus the electromagnetic momentum goes back to zero. 
 \end{itemize} 
 We see that the momentum is preserved over the temporal window over which the pulse is fully enclosed within the simulation window. The error on the conservation of the total momentum is slightly less than 1\%. It is reasonable in view of the chosen spatial and temporal steps. The numerical algorithm also preserves the conservation of momentum with a good accuracy.
 
In Fig. \ref{fig:Conservation}(b), we plot the linear density of energy as function of simulation time. 
The dashed red curve correspond to the electromagnetic energy density \cite{Jackson_1999}
\begin{equation}
U_\mathrm{em} = \frac{1}{2} \left[ \varepsilon_0 \textbf{E}^2 + \frac{1}{\mu_0}\textbf{B}^2  \right]
\end{equation}
that we have integrated over the $x-y$ plane. The dashed dotted blue (resp. dashed green) corresponds to the electron fluid (resp. ion fluid) energy density given by Eq. \eqref{Eq_Energy} and then integrated over the $x-y$ plane. The total energy plotted in black line is defined as the sum of electromagnetic, electron fluid and ion fluid energies. The linear density of energy of the input pulse in the $x-y$ plane can be calculated analytically and is given by: $E_\mathrm{Laser} = \frac{E_\mathrm{0}^2}{2} \sqrt{\frac{\varepsilon_\mathrm{0}}{\mu_\mathrm{0}}} \frac{3}{8} T w_0 \sqrt{\frac{\pi}{2}} = 0.19$~J/m. This analytical linear density of energy is shown as a black dotted line in Fig. \ref{fig:Conservation}(b).  

 We observe the three main sequences in Fig. \ref{fig:Conservation}(b):
 \begin{itemize}
     \item \textbf{1}: The electromagnetic energy increases as the pulse enters into the numerical window between $t=0$ and $t<50$~fs. At $t=50$~fs, the pulse is completely contained in the numerical window and has not yet interacted with the plasma ramp. The electromagnetic energy corresponds to the predicted analytical value $E_\mathrm{Laser}$.
     \item \textbf{2}: In the temporal window 70-130~fs, energy exchange with the plasma takes place (the electron energy increases).
     \item \textbf{3}: Between $170$~fs and $210$~fs, the reflected pulse leaves the integration volume and the electromagnetic energy decreases. No electromagnetic energy remains in the simulation window. This is expected for s-polarized wave. 
 \end{itemize} 
  We remark the conservation of the energy when the pulse is fully in the simulation window. The error on the conservation the total energy is around 0.1\%. The numerical algorithm also preserves the energy conservation with a good accuracy.

Figs. \ref{fig:Conservation}(a) and \ref{fig:Conservation}(b) demonstrated conservation of momentum and energy during s-polarized reflection over a plasma ramp.
 
\subsection*{Wave conversion on plasma density ramp}

In this second test, we consider the same numerical setup as shown in Fig. \ref{fig:Box}, but we inject a p-polarized laser pulse.

\begin{figure*}[h!]
\centering\includegraphics[scale=0.8]{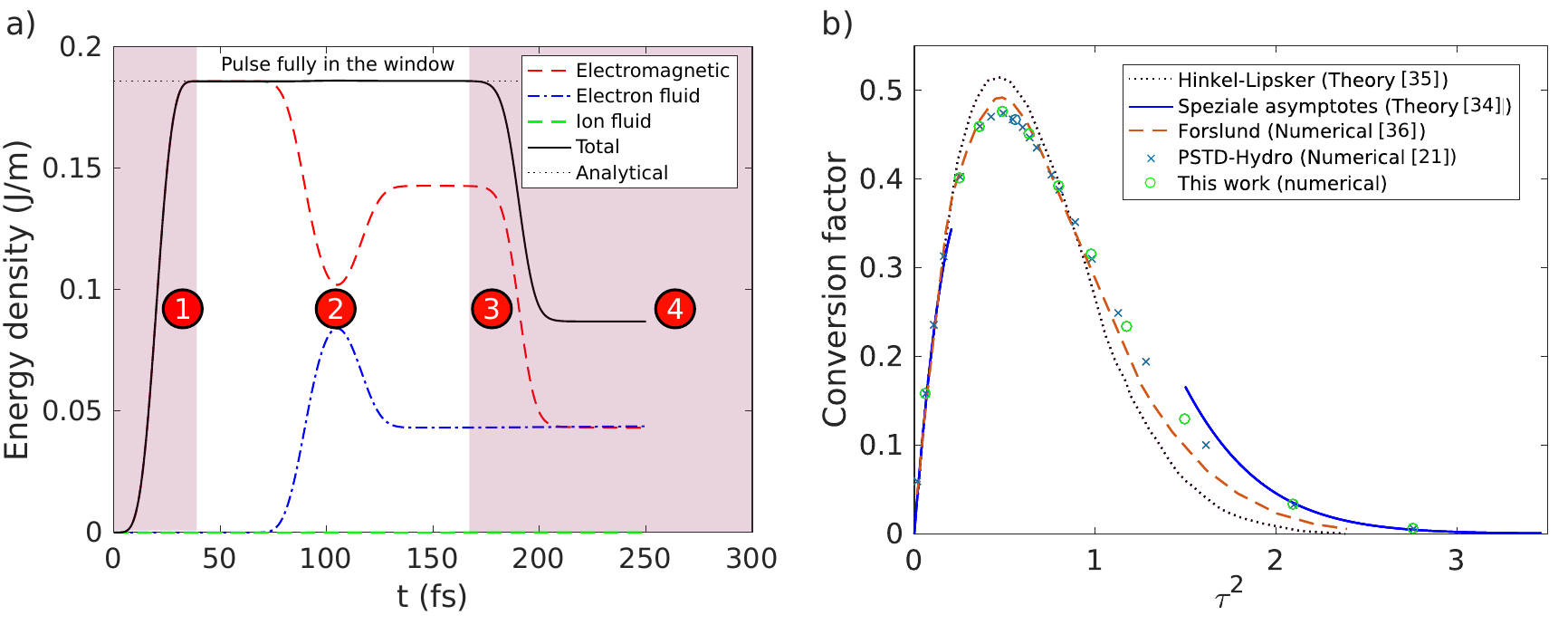}
\caption{(a) Evolution of electromagnetic energy, fluid energy, and total energy as function of time for p-polarized incident pulse. In the central white area, the pulse is fully enclosed within the numerical window and the conservation of the total energy is preserved. (b) Mode conversion factor as a function of $\tau^2 = \left(\frac{2 \pi L}{\lambda}\right)^{2/3}\sin^2\theta$. Green circles are our numerical results with PSATD/Hydrodynamic code. Results from others references are shown for comparison: Hinkel-Lipsker analytical model (dotted black line) \cite{hinkel-lipsker_analytic_1989}, Speziale asymptotes (solid blue lines) \cite{speziale_linear_1977}, Forslund PIC simulations\cite{forslund_theory_1975} (dashed orange line) and PSTD/Hydrodynamic (Blue crosses) \cite{Morel_2020}.}
\label{fig:AbsRes}
\end{figure*}

The energy of the system is plotted as a function of time in Fig. \ref{fig:AbsRes}(a). In the central white area, the error on the conservation the total energy is around 0.1\%. Furthermore, we observe for the sequence n$^\circ$4 that a fraction of the input energy remains in the simulation box while the laser pulse has left. This is due to the phenomenon of wave conversion onto a inhomogeneous plasma, {\it i.e.} conversion of an electromagnetic wave into a plasma wave which occur only for p-polarization \cite{speziale_linear_1977}. 

The conversion factor depends in particular on the plasma density gradient and the angle of incidence. Obtaining analytical solutions to this difficult problem usually requires a number of approximations. We performed a series of simulations with different angles of incidence, and we plot (blue circles) in Fig. \ref{fig:AbsRes}(b) the factor of the energy conversion as function of the quantity $\tau^2 = \left(\frac{2 \pi L}{\lambda}\right)^{2/3}\sin^2\theta$. 

 We compare conversion factors obtained with the PSATD/Hydrodynamic code (this work) and results of the literature. The PSATD/Hydro conversion factor curve is quantitatively superimposable to the one obtained the PSTD/hydro simulation (blue crosses) obtained in reference \cite{Morel_2020}. Our results are also in agreement with the analytical results of D.E. Hinkel-Lipsker {\it et al.} \cite{hinkel-lipsker_analytic_1989}, with T. Speziale {\it et al.} who described the asymptotic behaviors \cite{speziale_linear_1977} and also with those of D.W. Forslund {\it et al.} who have used Particle-In-Cell (PIC) simulations \cite{forslund_theory_1975}. The fact that we injected a short pulse (polychromatic) gaussian beam instead a monochromatic plane wave can explain the tenuous differences. In addition, the analytical results of references T. Speziale \textit{et al.}\cite{speziale_linear_1977} and D.E. Hinkel-Lipsker {\it et al.}\cite{hinkel-lipsker_analytic_1989}, have carried out assumptions that are not exactly fulfilled in our numerical test. This can explain the minor discrepancies observed. But overall, the results we obtained with the present solver are in good agreement with the state of the art.

\section*{Discussion}

In this section, we compare the benefits and drawbacks of PSATD/Hydro solver compared to PSTD/Hydro solver.
We numerically simulate a single cycle pulse plane wave in normal incidence onto a plasma ramp. The laser wavelength and plasma parameters are identical to the ones of Fig. \ref{fig:Box}. The pulse amplitude is $E_\mathrm{0}=4.3\times 10^{10}$~V/m. The computation is performed in 3D, with the same numerical sampling parameters as in  Fig. \ref{fig:Box}. We use the periodic boundary conditions in $y$ and $z$ directions. For the PSTD/Hydro solver, the time-step is fixed to $\Delta t =50$~as since we are constraint by Courant Friedrichs Lewy (CFL) conditions \cite{liu_pstd_1997}. In contrast, the time-step for the PSATD/Hydro solver is set to $\Delta t =200$~as, as it is only constrained by the sampling of the laser and the plasma wave frequencies.

\begin{figure*}[h!]
\centering\includegraphics[scale=0.8]{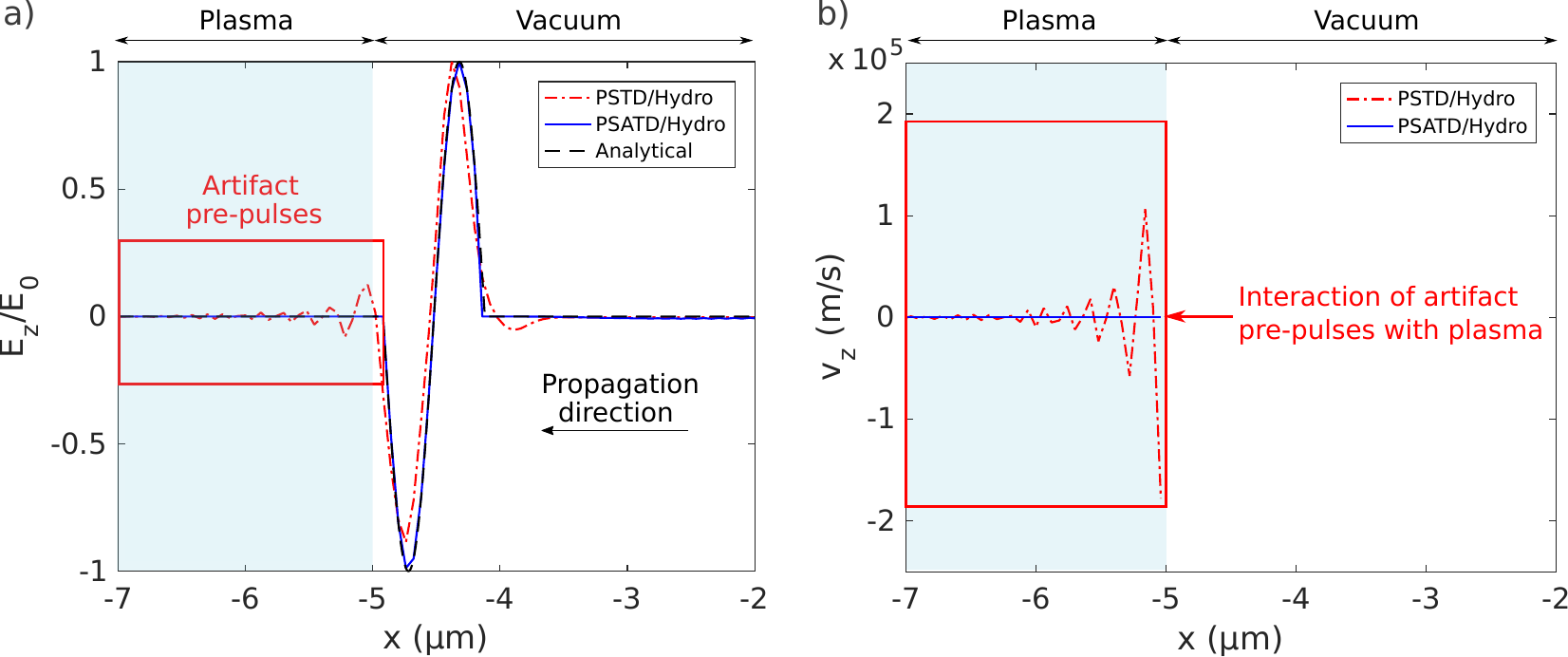}
\caption{(a) Single-cycle pulse propagation with PSTD/Hydro solver (red dashed-dotted line) and with PSATD/Hydro solver (solid blue line). The laser pulse interacts in normal incidence on a plasma ramp starting at $x \leq -5$~$\mu$m as shown in light blue on the figure. The PSATD/Hydro curve is superimposed on the analytical solution shown as a black dashed line. We observe artifact pre-pulses generated by numerical dispersion of PSTD in the plasma region. (b) Velocity component $v_\mathrm{z}$ of electron fluid at the same time.}
\label{fig:Veloc}
\end{figure*}

 In Fig. \ref{fig:Veloc}(a), we show a snapshot during propagation of the laser pulse with the different solvers. The snapshot is taken when the pulse has propagated through vacuum and just reaches the onset of the plasma ramp. We see that the PSATD/Hydro solver result (solid blue line) is precisely superimposed on the analytical solution in black dashed line. In contrast, the PSTD/Hydro solver (red dashed-dotted line) exhibits distortion of the laser pulse. Indeed, pre-pulses are generated by numerical dispersion of the PSTD algorithm. The amplitude of the last artifact pre-pulse (located at $x\approx -5$~$\mu$m) reaches around 15\% of the amplitude of the main peaks. 
 
  In Fig. \ref{fig:Veloc}(b), we plot the velocity component $v_\mathrm{z}$ of the electron fluid at the same time of the snapshot of Fig. \ref{fig:Veloc}(a). We observe that the laser pulse has not yet interacted with the plasma in the PSATD/Hydro simulation (blue line). However, in the PSTD/Hydro simulation (red dashed-dotted line), the artifact pre-pulses already interact with the plasma and accelerate the electrons to velocities around $10^5$~m/s. This effect is obviously undesirable, particularly in the case of the simulation of few cycle laser pulses with plasmas \cite{Mishra_2017}.

We also obtained better results in the PSATD/Hydro simulation whereas the time-step $\Delta t$ was 4 times greater than those in PSTD/Hydro simulation.

The PSATD/Hydro solver is well suited to pulse propagation. Specifically, the fact that PSATD is not constraint by the CFL condition, releases the strong numerical link between spatial and temporal sampling. The computational gain is therefore particularly significant in the case where high spatial resolution is required together with less demanding temporal resolution. We finish this section by reminding that the PSATD method requires that the background medium permittivity is uniform.

As a conclusion, we have developed a solver for the two-fluid plasma model based on a relatively simple technique, which does not necessitate staggered grids and which benefits of fundamentally having no numerical dispersion for the propagation of electromagnetic waves in absence of current. The algorithm relies on the PseudoSpectral Analytical Time-Domain (PSATD) technique which is a powerful method for propagating laser pulses, and on a combination of two-step Lax-Wendroff (LW) and fourth-order Runge-Kutta (RK4) for the fluid equations. We have demonstrated that the PSATD/Hydro solver preserves momentum and energy during a test with s-polarized laser pulse incident over a plasma ramp. The tests of wave conversion on plasma ramps have demonstrated an excellent quantitative agreement with numerical and analytical results of the state of the art. We have shown that PSATD/Hydro solver has two main advantages compared to the PSTD/Hydro solver: the pre-pulses generated by numerical dispersion are removed and the time-step is not constraint by CFL conditions. For simulations which require low temporal resolution and high spatial resolution, the gain in terms of computational resources with PSATD/Hydro solver can be really significant. The PSATD/Hydro solver is a computationally inexpensive but powerful tool for the study of laser-plasma interaction.

\section*{Methods}

Simulations were performed with Nvidia Tesla K40 GPU card. This card has 12~GB memory size, 2880 CUDA cores and 745 MHz processor core clock.

\section*{Acknowledgements}

Numerical simulations have been performed using the M\'{e}socentre de Calcul de Franche-Comt\'{e}. The research leading to these results has received funding from the European Research Council (ERC) under the European Union's Horizon 2020 research and innovation program (grant agreement No 682032-PULSAR), R\'egion Bourgogne Franche-Comt\'e, the EIPHI Graduate School (ANR-17-EURE-0002) and I-SITE BFC project (ANR-15-IDEX-0003).

\section*{Author contributions statement}

B.M. conceived the algorithm structure, implemented the fluid solver and performed the simulations, R.G. implemented the PSATD solver, K.A. and F.C. analysed the results. B.M. and F.C. prepared the manuscript which was reviewed by all authors. 
\bibliographystyle{naturemag}
\bibliography{MyBib}

\begin{thebibliography}{10}
\expandafter\ifx\csname url\endcsname\relax
  \def\url#1{\texttt{#1}}\fi
\expandafter\ifx\csname urlprefix\endcsname\relax\def\urlprefix{URL }\fi
\providecommand{\bibinfo}[2]{#2}
\providecommand{\eprint}[2][]{\url{#2}}

\bibitem{Strickland_1985}
\bibinfo{author}{Strickland, D.} \& \bibinfo{author}{Mourou, G.}
\newblock \bibinfo{title}{Compression of amplified chirped optical pulses}.
\newblock \emph{\bibinfo{journal}{Opt. Commun.}} \textbf{\bibinfo{volume}{55}},
  \bibinfo{pages}{447--449} (\bibinfo{year}{1985}).
\newblock
  \urlprefix\url{http://www.sciencedirect.com/science/article/pii/0030401885901518}.

\bibitem{eliezer_applications_2008}
\bibinfo{author}{Eliezer, S.} \& \bibinfo{author}{Mima, K.}
\newblock \emph{\bibinfo{title}{Applications of {Laser}-{Plasma}
  {Interactions}}} (\bibinfo{publisher}{CRC Press}, \bibinfo{year}{2008}).

\bibitem{Basov_1964}
\bibinfo{author}{Basov, N.} \& \bibinfo{author}{Krokhin, O.}
\newblock \bibinfo{title}{Condition for heating up of a plasma by the radiation
  from an optical generator}.
\newblock \emph{\bibinfo{journal}{J. Exp. Theor. Phys}}
  \textbf{\bibinfo{volume}{19}}, \bibinfo{pages}{123--125}
  (\bibinfo{year}{1964}).

\bibitem{Esarey_1996}
\bibinfo{author}{{Esarey}, E.}, \bibinfo{author}{{Sprangle}, P.},
  \bibinfo{author}{{Krall}, J.} \& \bibinfo{author}{{Ting}, A.}
\newblock \bibinfo{title}{Overview of plasma-based accelerator concepts}.
\newblock \emph{\bibinfo{journal}{IEEE Plasma Sci}}
  \textbf{\bibinfo{volume}{24}}, \bibinfo{pages}{252--288}
  (\bibinfo{year}{1996}).

\bibitem{Daido_2002}
\bibinfo{author}{Daido, H.}
\newblock \bibinfo{title}{Review of soft x-ray laser researches and
  developments}.
\newblock \emph{\bibinfo{journal}{Rep. Prog. Phys.}}
  \textbf{\bibinfo{volume}{65}}, \bibinfo{pages}{1513--1576}
  (\bibinfo{year}{2002}).

\bibitem{Kauranen_2012}
\bibinfo{author}{Kauranen, M.} \& \bibinfo{author}{Zayats, A.}
\newblock \bibinfo{title}{Nonlinear plasmonics}.
\newblock \emph{\bibinfo{journal}{Nat. Photonics}}
  \textbf{\bibinfo{volume}{6}}, \bibinfo{pages}{737–748}
  (\bibinfo{year}{2012}).

\bibitem{mckenna_laser-plasma_nodate}
\bibinfo{author}{McKenna, P.}, \bibinfo{author}{Neely, D.},
  \bibinfo{author}{Bingham, R.} \& \bibinfo{author}{Jaroszynski, D.}
\newblock \emph{\bibinfo{title}{Laser-{Plasma} {Interactions} and
  {Applications}}} (\bibinfo{publisher}{Springer}, \bibinfo{year}{2013}).

\bibitem{gibbon_short_2005}
\bibinfo{author}{Gibbon, P.}
\newblock \emph{\bibinfo{title}{Short {Pulse} {Laser} {Interactions} with
  {Matter}: {An} {Introduction}}} (\bibinfo{publisher}{Imperial College Press},
  \bibinfo{year}{2005}).

\bibitem{abgrall_robust_2014}
\bibinfo{author}{Abgrall, R.} \& \bibinfo{author}{Kumar, H.}
\newblock \bibinfo{title}{Robust {Finite} {Volume} {Schemes} for {Two}-{Fluid}
  {Plasma} {Equations}}.
\newblock \emph{\bibinfo{journal}{J. Sci. Comput.}}
  \textbf{\bibinfo{volume}{60}}, \bibinfo{pages}{584--611}
  (\bibinfo{year}{2014}).

\bibitem{shumlak_approximate_2003}
\bibinfo{author}{Shumlak, U.} \& \bibinfo{author}{Loverich, J.}
\newblock \bibinfo{title}{Approximate {Riemann} solver for the two-fluid plasma
  model}.
\newblock \emph{\bibinfo{journal}{J. Comput. Phys.}}
  \textbf{\bibinfo{volume}{187}}, \bibinfo{pages}{620--638}
  (\bibinfo{year}{2003}).

\bibitem{roe_approximate_1981}
\bibinfo{author}{Roe, P.~L.}
\newblock \bibinfo{title}{Approximate {Riemann} solvers, parameter vectors, and
  difference schemes}.
\newblock \emph{\bibinfo{journal}{J. Comput. Phys.}}
  \textbf{\bibinfo{volume}{43}}, \bibinfo{pages}{357--372}
  (\bibinfo{year}{1981}).

\bibitem{loverich_discontinuous_2005}
\bibinfo{author}{Loverich, J.} \& \bibinfo{author}{Shumlak, U.}
\newblock \bibinfo{title}{A discontinuous {Galerkin} method for the full
  two-fluid plasma model}.
\newblock \emph{\bibinfo{journal}{Comput. Phys. Commun.}}
  \textbf{\bibinfo{volume}{169}}, \bibinfo{pages}{251--255}
  (\bibinfo{year}{2005}).

\bibitem{loverich_discontinuous_2011}
\bibinfo{author}{Loverich, J.}, \bibinfo{author}{Hakim, A.} \&
  \bibinfo{author}{Shumlak, U.}
\newblock \bibinfo{title}{A {Discontinuous} {Galerkin} {Method} for {Ideal}
  {Two}-{Fluid} {Plasma} {Equations}}.
\newblock \emph{\bibinfo{journal}{Commun. Comput. Phys.}}
  \textbf{\bibinfo{volume}{9}}, \bibinfo{pages}{240--268}
  (\bibinfo{year}{2011}).

\bibitem{srinivasan_analytical_2011}
\bibinfo{author}{Srinivasan, B.} \& \bibinfo{author}{Shumlak, U.}
\newblock \bibinfo{title}{Analytical and computational study of the ideal full
  two-fluid plasma model and asymptotic approximations for
  {Hall}-magnetohydrodynamics}.
\newblock \emph{\bibinfo{journal}{Phys. Plasmas}}
  \textbf{\bibinfo{volume}{18}}, \bibinfo{pages}{092--113}
  (\bibinfo{year}{2011}).

\bibitem{sousa_blended_2016}
\bibinfo{author}{Sousa, E.~M.} \& \bibinfo{author}{Shumlak, U.}
\newblock \bibinfo{title}{A blended continuous-discontinuous finite element
  method for solving the multi-fluid plasma model}.
\newblock \emph{\bibinfo{journal}{J. Comput. Phys.}}
  \textbf{\bibinfo{volume}{326}}, \bibinfo{pages}{56--75}
  (\bibinfo{year}{2016}).

\bibitem{alvarez_laguna_fully-implicit_2018}
\bibinfo{author}{Alvarez~Laguna, A.}, \bibinfo{author}{Ozak, N.},
  \bibinfo{author}{Lani, A.}, \bibinfo{author}{Deconinck, H.} \&
  \bibinfo{author}{Poedts, S.}
\newblock \bibinfo{title}{Fully-implicit finite volume method for the ideal
  two-fluid plasma model}.
\newblock \emph{\bibinfo{journal}{Comput. Phys. Commun.}}
  \textbf{\bibinfo{volume}{231}}, \bibinfo{pages}{31--44}
  (\bibinfo{year}{2018}).

\bibitem{mason_electromagnetic_1987}
\bibinfo{author}{Mason, R.~J.}
\newblock \bibinfo{title}{An electromagnetic field algorithm for 2{D} implicit
  plasma simulation}.
\newblock \emph{\bibinfo{journal}{J. Comput. Phys.}}
  \textbf{\bibinfo{volume}{71}} (\bibinfo{year}{1987}).

\bibitem{mason_hybrid_1986}
\bibinfo{author}{Mason, R.~J.} \& \bibinfo{author}{Cranfill, C.~W.}
\newblock \bibinfo{title}{Hybrid {Two}-{Dimensional} {Electron} {Transport} in
  {Self}-{Consistent} {Electromagnetic} {Fields}}.
\newblock \emph{\bibinfo{journal}{IEEE T. Plasma Sci.}}
  \textbf{\bibinfo{volume}{14}}, \bibinfo{pages}{45--52}
  (\bibinfo{year}{1986}).

\bibitem{baboolal_two-scale_2007}
\bibinfo{author}{Baboolal, S.} \& \bibinfo{author}{Bharuthram, R.}
\newblock \bibinfo{title}{Two-scale {Numerical} {Solution} of the
  {Electromagnetic} {Two}-fluid plasma-{Maxwell} {Equations}: {Shock} and
  {Soliton} {Simulation}}.
\newblock \emph{\bibinfo{journal}{Math. Comput. Simul.}}
  \textbf{\bibinfo{volume}{76}}, \bibinfo{pages}{3--7} (\bibinfo{year}{2007}).

\bibitem{kumar_entropy_2012}
\bibinfo{author}{Kumar, H.} \& \bibinfo{author}{Mishra, S.}
\newblock \bibinfo{title}{Entropy {Stable} {Numerical} {Schemes} for
  {Two}-{Fluid} {Plasma} {Equations}}.
\newblock \emph{\bibinfo{journal}{J. Sci. Comput.}}
  \textbf{\bibinfo{volume}{52}}, \bibinfo{pages}{401--425}
  (\bibinfo{year}{2012}).

\bibitem{Morel_2020}
\bibinfo{author}{Morel, B.}, \bibinfo{author}{Giust, R.},
  \bibinfo{author}{Ardaneh, K.} \& \bibinfo{author}{Courvoisier, F.}
\newblock \bibinfo{title}{A simple solver for the two-fluid plasma model based
  on {PseudoSpectral} time-domain algorithm}.
\newblock \emph{\bibinfo{journal}{Comm. Comput. Phys.}}
  \textbf{\bibinfo{volume}{29}}, \bibinfo{pages}{955--978}
  (\bibinfo{year}{2021}).

\bibitem{liu_pstd_1997}
\bibinfo{author}{Liu, Q.~H.}
\newblock \bibinfo{title}{The {PSTD} algorithm: {A} time-domain method
  requiring only two cells per wavelength}.
\newblock \emph{\bibinfo{journal}{Microw. Opt. Technol. Lett.}}
  \textbf{\bibinfo{volume}{15}}, \bibinfo{pages}{158--165}
  (\bibinfo{year}{1997}).

\bibitem{liska_composite_1998}
\bibinfo{author}{Liska, R.} \& \bibinfo{author}{Wendroff, B.}
\newblock \bibinfo{title}{Composite {Schemes} for {Conservation} {Laws}}.
\newblock \emph{\bibinfo{journal}{SIAM J. Numer. Anal.}}
  \textbf{\bibinfo{volume}{35}}, \bibinfo{pages}{2250--2271}
  (\bibinfo{year}{1998}).

\bibitem{Vay_2013}
\bibinfo{author}{Vay, J.-L.}, \bibinfo{author}{Haber, I.} \&
  \bibinfo{author}{Godfrey, B.~B.}
\newblock \bibinfo{title}{A domain decomposition method for pseudo-spectral
  electromagnetic simulations of plasmas}.
\newblock \emph{\bibinfo{journal}{J. Comput. Phys.}}
  \textbf{\bibinfo{volume}{243}}, \bibinfo{pages}{260--268}
  (\bibinfo{year}{2013}).
\newblock
  \urlprefix\url{http://www.sciencedirect.com/science/article/pii/S0021999113001873}.

\bibitem{strang_construction_1968}
\bibinfo{author}{Strang, G.}
\newblock \bibinfo{title}{On the {Construction} and {Comparison} of
  {Difference} {Schemes}}.
\newblock \emph{\bibinfo{journal}{SIAM J. Numer. Anal.}}
  \textbf{\bibinfo{volume}{5}}, \bibinfo{pages}{506--517}
  (\bibinfo{year}{1968}).

\bibitem{a._hakim_high_2006}
\bibinfo{author}{Hakim, A.}, \bibinfo{author}{Loverich, J.} \&
  \bibinfo{author}{Shumlak, U.}
\newblock \bibinfo{title}{A high resolution wave propagation scheme for ideal
  {Two}-{Fluid} plasma equations}.
\newblock \emph{\bibinfo{journal}{J. Comput. Phys.}}
  \textbf{\bibinfo{volume}{219}}, \bibinfo{pages}{418--442}
  (\bibinfo{year}{2006}).

\bibitem{bittencourt_fundamentals_2013}
\bibinfo{author}{Bittencourt, J.~A.}
\newblock \emph{\bibinfo{title}{Fundamentals of {Plasma} {Physics}}}
  (\bibinfo{publisher}{Springer}, \bibinfo{year}{2013}).

\bibitem{yee_numerical_1966}
\bibinfo{author}{Yee, K.}
\newblock \bibinfo{title}{Numerical solution of initial boundary value problems
  involving {Maxwell's} equations in isotropic media}.
\newblock \emph{\bibinfo{journal}{IEEE Trans. Antennas Propag.}}
  \textbf{\bibinfo{volume}{14}}, \bibinfo{pages}{302--307}
  (\bibinfo{year}{1966}).

\bibitem{Shapoval_2019}
\bibinfo{author}{Shapoval, O.}, \bibinfo{author}{Vay, J.-L.} \&
  \bibinfo{author}{Vincenti, H.}
\newblock \bibinfo{title}{Two-step perfectly matched layer for arbitrary-order
  pseudo-spectral analytical time-domain methods}.
\newblock \emph{\bibinfo{journal}{Comput. Phys. Commun.}}
  \textbf{\bibinfo{volume}{235}}, \bibinfo{pages}{102--110}
  (\bibinfo{year}{2019}).
\newblock
  \urlprefix\url{http://www.sciencedirect.com/science/article/pii/S0010465518303291}.

\bibitem{richtmyer_survey_1962}
\bibinfo{author}{Richtmyer, D.}
\newblock \emph{\bibinfo{title}{A {Survey} of {Difference} {Methods} for
  {Non}-{Steady} {Fluid} {Dynamics}}} (\bibinfo{publisher}{Technical Note
  NCAR/TN-63-2+STR}, \bibinfo{year}{1962}).

\bibitem{shampine_two-step_2005}
\bibinfo{author}{Shampine, L.~F.}
\newblock \bibinfo{title}{Two-step {Lax}–{Friedrichs} method}.
\newblock \emph{\bibinfo{journal}{Appl. Math. Lett.}}
  \textbf{\bibinfo{volume}{18}}, \bibinfo{pages}{1134--1136}
  (\bibinfo{year}{2005}).

\bibitem{leveque_finite_2002}
\bibinfo{author}{Leveque, R.~J.}
\newblock \emph{\bibinfo{title}{Finite {Volume} {Methods} for {Hyperbolic}
  {Problems}}} (\bibinfo{publisher}{Cambridge University Press},
  \bibinfo{year}{2002}).

\bibitem{Jackson_1999}
\bibinfo{author}{Jackson, J.~D.}
\newblock \emph{\bibinfo{title}{Classical electrodynamics}}
  (\bibinfo{publisher}{John Wiley \& Sons}, \bibinfo{year}{1998}).
\newblock \bibinfo{note}{Third edition}.

\bibitem{speziale_linear_1977}
\bibinfo{author}{Speziale, T.} \& \bibinfo{author}{Catto, P.~J.}
\newblock \bibinfo{title}{Linear wave conversion in an unmagnetized,
  collisionless plasma}.
\newblock \emph{\bibinfo{journal}{Phys. Fluids}} \textbf{\bibinfo{volume}{20}},
  \bibinfo{pages}{990--997} (\bibinfo{year}{1977}).

\bibitem{hinkel-lipsker_analytic_1989}
\bibinfo{author}{Hinkel-Lipsker, D.~E.}, \bibinfo{author}{Fried, B.~D.} \&
  \bibinfo{author}{Morales, G.~J.}
\newblock \bibinfo{title}{Analytic expression for mode conversion of {Langmuir}
  and electromagnetic waves}.
\newblock \emph{\bibinfo{journal}{Phys. Rev. Lett.}}
  \textbf{\bibinfo{volume}{62}}, \bibinfo{pages}{2680--2682}
  (\bibinfo{year}{1989}).

\bibitem{forslund_theory_1975}
\bibinfo{author}{Forslund, D.~W.}, \bibinfo{author}{Kindel, J.~M.},
  \bibinfo{author}{Lee, K.}, \bibinfo{author}{Lindman, E.~L.} \&
  \bibinfo{author}{Morse, R.~L.}
\newblock \bibinfo{title}{Theory and simulation of resonant absorption in a hot
  plasma}.
\newblock \emph{\bibinfo{journal}{Phys. Rev. A}} \textbf{\bibinfo{volume}{11}},
  \bibinfo{pages}{679--683} (\bibinfo{year}{1975}).

\bibitem{Mishra_2017}
\bibinfo{author}{Mishra, S.~K.}, \bibinfo{author}{Andreev, A.} \&
  \bibinfo{author}{Kalashinikov, M.~P.}
\newblock \bibinfo{title}{Reflection of few cycle laser pulses from an
  inhomogeneous overdense plasma}.
\newblock \emph{\bibinfo{journal}{Opt. Express}} \textbf{\bibinfo{volume}{25}},
  \bibinfo{pages}{11637--11651} (\bibinfo{year}{2017}).

\end{thebibliography}

\end{document}